# Chersiphron & Son Engineers

**Amelia Carolina Sparavigna**
**Dipartimento di Fisica**
**Politecnico di Torino**

An ancient engineering firm worked successfully in the construction of one of the Seven Wonders of the Ancient World. The engineers used inclined planes, bags of sand and shafts of columns and architraves as wheels and axels.

Chersiphron and his son Metagenes were among the engineers and architects who worked in the creation of one of the Seven Wonders of the Ancient World, the Temple of Artemis at Ephesus [1]. We find in the books written by Pliny the Elder and Vitruvius, where Chersiphron is recalled as Ctesiphon, some interesting descriptions on how the architect worked and solved some problems of engineering. Let us remember that Ephesus was a Greek town in Asia Minor, located near the modern town of Selçuk in Turkey. Nowadays, only foundations and some sculptural elements of the latest temple can be seen at the archaeological site. Figure 1 shows the satellite map of the remains of old Ephesus and its temple. During the Hellenistic period, the local river silted up the ancient harbor. As we can see in the satellite images, it is far from the sea of approximately six kilometers. Later, Ephesus became a major Roman town. It was probably among the largest cities in the ancient world.

During its history, the temple of Artemis was completely rebuilt several times before its final destruction [2,3]. The site of the temple was first occupied by some sanctuaries of the Bronze Age. In the seventh century BC, the place was destroyed by a flood. The Ephesians decided its reconstruction around 550 BC, calling the Cretan architect and engineer Chersiphron and his son Metagenes. The work took several years. They used marble to build the temple, with columns that stood in double rows to form a ceremonial passage around the inner chamber housing the goddess' statue.

Both Pliny the Elder and Vitruvius report some discussion on the building of the temple and on the technologies used during the work. Let us start with what Pliny is telling in his Natural History [4].

> *"A marshy soil was selected for its site, in order that it might not suffer from earthquakes, or the chasms which they produce. On the other hand, again, that the foundations of so vast a pile might not have to rest upon a loose and shifting bed, layers of trodden charcoal were placed beneath, with fleeces covered with wool upon the top of them. The entire length of the temple is four hundred and twenty-five feet, and the breadth two hundred and twenty-five."*

Pliny seems to ignore that the site of the temple was more ancient and that the architects were rebuilding on it, therefore he guessed that a marshy soil was chosen for taking precaution against earthquakes. Unfortunately, this is not true, as the 1985 Mexico City earthquake had demonstrated. The epicenter was near Lázaro Cárdenas, 350 km away, but "in the marshy soil underlying the (Mexico) city centre, the destructive force of the earthquake matched that at the epicenter." [5] Another problem of marshy lands can be the phenomenon of soil liquefaction. The soil, saturated by water, loses its strength during the earthquake shaking and behaves like a liquid [6]. The Latin naturalist describes that the architects obtained the consolidation of soil, necessary to have a proper basement for the temple, using some trodden charcoal. It would be interesting to check the mechanical properties of such a substrate, with respect to the seismic waves and soil liquefaction.

Pliny is also telling that a layer of fleeces was used to cover the substrate of charcoal. We can try to guess a reason for the use of these fleeces: probably, the architects ordered to prepare this layer to have a sort of seismic isolation of the temple from the ground, allowing it to float on the soil during the earthquakes as a boat on water. This is, more or less, the same behavior of a modern anti-seismic building.

The anti-seismic technology consists of installation of some isolators, which decouple the buildings from the ground [7]. Such isolators are cylinders consisting of alternate layers of rubber and steel bonded together, with a central lead core. The rubber layers allow the isolator to displace sideways, because they are very soft. However, the structure is very stiff vertically, because the layers of rubber reinforced by steel. These two characteristics allow the isolator to move laterally, and, at the same time, they can carry significant axial load. The lead core acts as a damper of oscillations. Pliny continues his description

> "Ctesiphon was the architect who presided over the work. The great marvel in this building is, how such ponderous architraves (epistylia) could possibly have been raised to so great a height. This, however, the architect effected by means of bags filled with sand, which he piled up upon an inclined plane until they reached beyond the capitals of the columns; then, as he gradually emptied the lower bags, the architraves insensibly settled in the places assigned them."

In building the temple, people used the inclined planes to enable an easy work against gravity of huge masses. It is remarkable the use of bags filled with sand to set the huge stones. It is highly probable that ancient Egyptians had developed several techniques based on the use of sand to put stones and obelisks in their proper positions and that this technology had migrated within the Mediterranean region. The temple built by Chersiphron and Metagenes was destroyed by fire on 356 BC. The Ephesians rebuilt it, starting from 323 BC, larger and with more than 127 columns. The temple was so beautiful (we can imagine it as in Figure 2), that Antipater of Sidon, a poet who lived in the 2nd century BC, when compiled his list of the Seven Wonders, told [8]:

> "I have set eyes on the wall of lofty Babylon on which is a road for chariots, and the statue of Zeus by the Alpheus, and the hanging gardens, and the colossus of the Sun, and the huge labour of the high pyramids, and the vast tomb of Mausolus; but when I saw the house of Artemis that mounted to the clouds, those other marvels lost their brilliancy, and I said, "Lo, apart from Olympus, the Sun never looked on aught so grand."

Chersiphron then was co-author of the building of the marble temple with his son Metagenes. Besides the use of inclined planes and sand to set the elements of the temple, these engineers devised two interesting methods to move columns and architraves. Vitruvius in his "De Architecture" explains "the ingenious contrivance of Ctesiphon" to move the columns [9].

> "When he removed from the quarry the shafts of the columns, ... not thinking it prudent to trust them on carriages, lest their weight should sink the wheels in the

> *soft roads over which they would have to pass, he devised the following scheme. He made a frame of four pieces of timber, two of which were equal in length to the shafts of the columns, and were held together by the two transverse. In each end of the shaft he inserted iron pivots, whose ends were dovetailed thereinto, and run with lead. The pivots worked in gudgeons fastened to the timber frame, whereto were attached oaken shafts. The pivots having a free revolution in the gudgeons, when the oxen were attached and drew the frame, the shafts rolled round, and might have been conveyed to any distance.*"

We can see how shafts were moved in Ref.[10]. Figure 3 shows the wooden frame used to move the marble cylinders. It is like a modern roller compactor, used to smooth the surface of roads: at those times, oxen were pulling it instead of engines. We can also tell that after the conveyance of the first shafts, the soil of the road to the temple probably became more compact, therefore increasing the ease of movement of the following stones. The problem is that we have also stone entablatures and architraves to move. This was the job of Megatenes [9].

> "*The shafts having been thus transported, the entablatures were to be removed, when Metagenes the son of Ctesiphon, applied the principle upon which the shafts had been conveyed to the removal of those also. He constructed wheels about twelve feet diameter, and fixed the ends of the blocks of stone whereof the entablature was composed into them; pivots and gudgeons were then prepared to receive them in the manner just described, so that when the oxen drew the machine, the pivots turning in the gudgeons, caused the wheels to revolve, and thus the blocks, being enclosed like axles in the wheels, were brought to the work without delay, as were the shafts of the columns. … But the method would not have been practicable for any considerable distance. From the quarries to the temple is a length of not more than eight thousand feet, and the interval is a plain without any declivity.*"

Again, we can see how entablatures were moved in Ref.[10] and in the Figure 4. The idea of Chersiphron was essentially to use column shafts as wheels. A further development of this method allowed Metagenes to use architraves as axles, around whose ends he prepared wheels of wood.

The two engineers were so successful to become a model in ancient times, as the discussion of their works in the books by Pliny and Vitruvius demonstrate. What is remarkable is their skills in developing machines suitable for the local necessities, for instance, the conveyance of huge stone form the quarries to the site of the temple on a soft soil. It is also interesting the use of inclined planes and bags of sand as the scaffolding of the temple. It is possible to guess that this scaffolding gradually increased during the construction, reaching the roof of the temple and helping in setting columns and architraves. Then it had been gradually removed, starting from the top of the temple: meanwhile, statues and other decorations were set during the scaffolding dismissal. Probably the construction of this temple was safer and more secure than some constructions of present days.

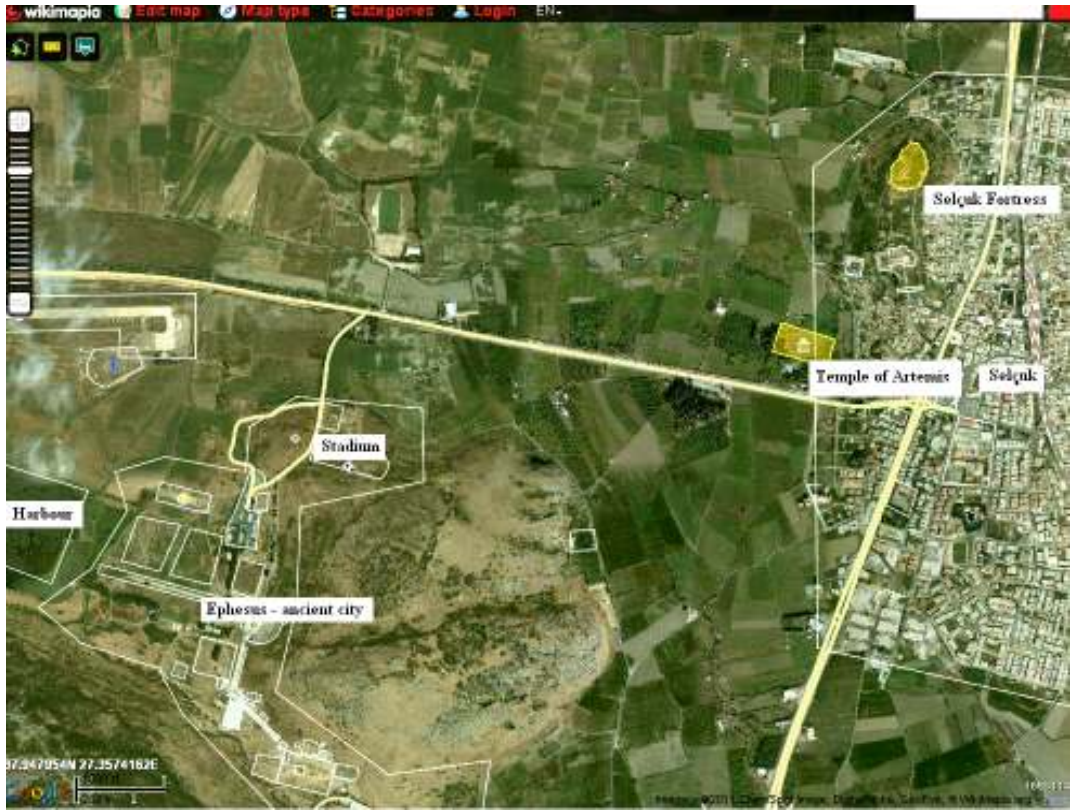

Fig.1: Satellite map from Wikimapia of old Ephesus and the location of the temple. During the Hellenistic period, the local river silted up the ancient harbor. It is now far from the sea of approximately six kilometers.

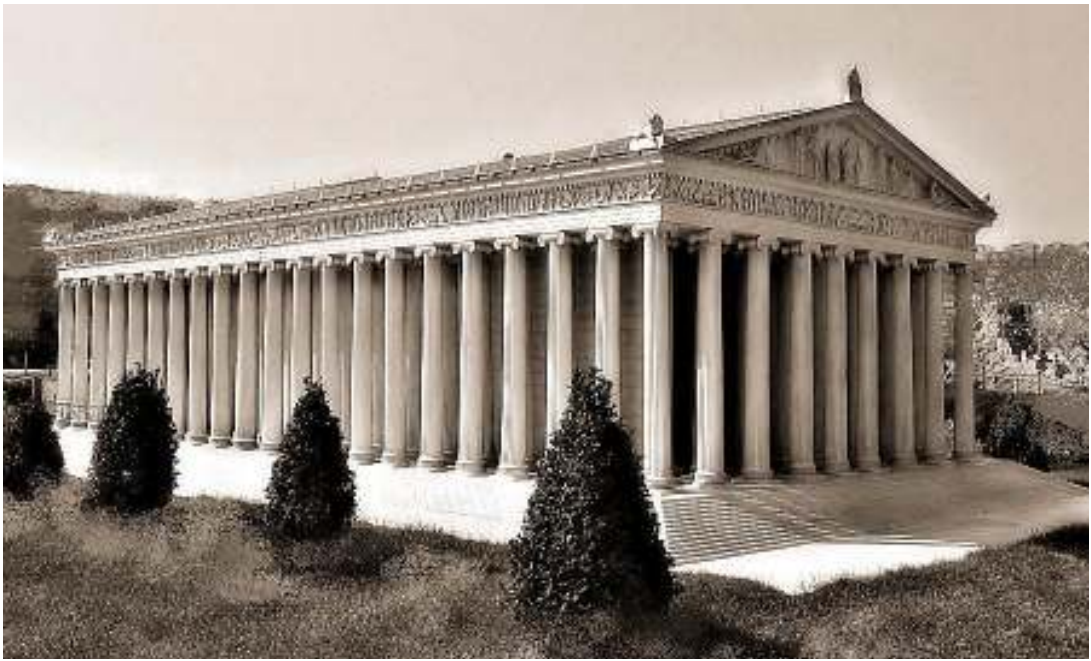

Fig.2: The Temple of Artemis. Image obtained from a picture of a model of the temple (credits: Zee Prime at cs.wikipedia).

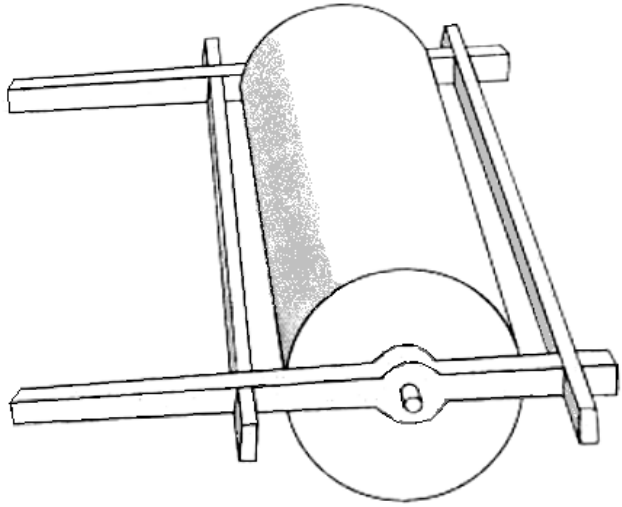

Fig.3: The Chersiphron's machine to move shafts of columns. The marble cylinder is framed by four timbers, and it can revolve about pivots. This machine is like the modern roller compactors of roads.

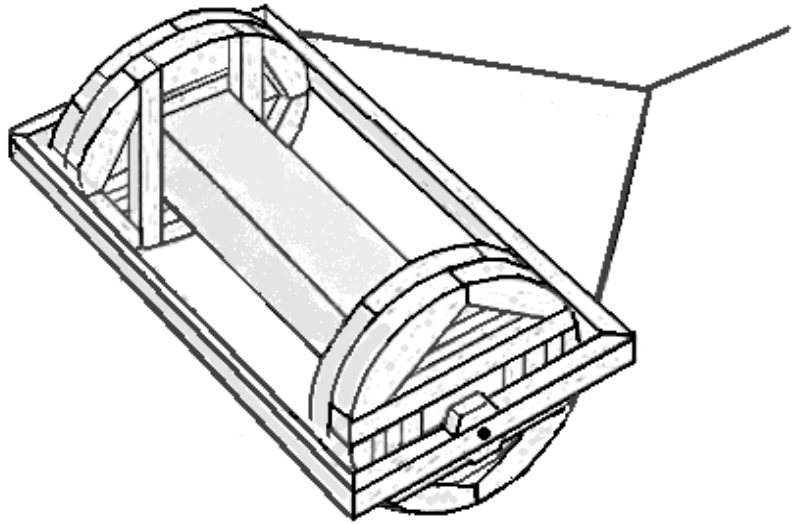

Fig.4: The Megatenes' machine to move architraves. The wooden wheels were enclosed to the architraves, which in this way were forming the axles of the machine.